\documentclass[11pt]{JHEP3}
\pdfoutput=1 

\usepackage{amsmath,amssymb}
\usepackage[T1]{fontenc} 
\usepackage{slashed}
\newcommand{\be}{\begin{equation}}
\newcommand{\ee}{\end{equation}}
\newcommand{\ben}{\begin{eqnarray}}
\newcommand{\een}{\end{eqnarray}}

\newcommand{\ads}{{\text{AdS}_5}}

\newcommand{\refb}[1]{(\ref{#1})}

\def\one{{\hbox{ 1\kern-.8mm l}}}
\def\zero{{\hbox{ 0\kern-1.5mm 0}}}

\def\cn{{\mathcal N}}

\def\tr{{\rm Tr}}
\def\sl{\slashed}
\def\l{\left}
\def\r{\right}

\def\s{\sigma}

\def\o[#1]{{\rm O}\left({#1}\right)}
\def\dotl[#1,#2]{\left\langle #1, #2 \right\rangle}
\def\dotlb[#1,#2]{[ #1, #2 ]}
\def\dotp[#1,#2]{(#1) \cdot (#2)}
\def\>{\rangle}
\def\<{\langle}

\def\ss2{\text{S}^2}

\def\adss{\text{AdS}_2}

\def\s2s2{\ss2\otimes\ss2}
\def\ads2s2{\adss\otimes\ss2}
\def\s2s2n{\l(\ss2\otimes\ss2\r)/\mathbb{Z}_N}
\def\ads2s2n{\l(\adss\otimes\ss2\r)/\mathbb{Z}_N}
\def\zn{\mathbb{Z}_N}

\def\bs{\bar{s}}

\def\ot{\mathcal{O}\l(t\r)}

\title{Logarithmic Corrections to Twisted Indices from the Quantum Entropy Function}

\author{Abhishek Chowdhury$^a$\footnote{abhishek AT hri DOT res DOT in}, Rajesh Kumar Gupta$^b$\footnote{ rgupta AT ictp DOT it}, Shailesh Lal$^c$\footnote{shailesh DOT hri AT gmail DOT com}, Milind Shyani$^{a,d}$\footnote{milindshyani AT gmail DOT com}, Somyadip Thakur$^e$\footnote{somyadip AT cts DOT iisc DOT ernet DOT in}\\

$^a$ $\,$Harish-Chandra Research Institute,\\
$\;$ $\,$Chhatnag Road, Jhusi, Allahabad -- 211019, India.\\

$^b$ $\,$ICTP, High Energy, Cosmology and Astroparticle Physics,\\
$\;$ $\,$Strada Costiera 11, 34151, Trieste, Italy.\\

$^c$ $\,$Department of Physics and Astronomy, \\
$\;$ $\,$Seoul National University \\
$\;$ $\,$Seoul 151-747, Korea. \\

$^d$ $\,$Birla Institute of Technology and Science -- Pilani,\\
$\;$ $\,$K.K. Birla Goa Campus, India.\\

$^e$ $\,$Centre for High Energy Physics,\\
$\;$ $\,$Indian Institute of Science, C.V. Raman Avenue,\\
$\;$ $\,$Bangalore 560012, India.\\
}

\abstract{We compute logarithmic corrections to the twisted index $B^g_6$ in four-dimensional $\cn=4$ and $\cn=8$ string theories using the framework of the Quantum Entropy Function. We find that these vanish, matching perfectly with the large--charge expansion of the corresponding microscopic expressions.}

\preprint{HRI/ST1404\\SNUTP14-004}
\begin{document} 

\section{Introduction and Review}
Indices carry important information about the spectrum of dyons in string theory. In particular, in four dimensional string theories the helicity trace index, defined by \cite{Bachas:1996bp,Gregori:1997hi}
\begin{equation}
B_{2n} = {1\over \l(2n\r)!}\tr\l[\l(-1\r)^{2h}\l(2h\r)^{2n}\r]
\end{equation}
receives contributions only from those BPS states in the string theory which break less than $4n$ supersymmetries. Here the trace is over all states in the string theory that carry some specified electric and magnetic charges. This has now been computed exactly for a wide class of $\cn=4$ and $\cn=8$ string theories \cite{Dijkgraaf:1996it,Maldacena:1999bp,LopesCardoso:2004xf,Shih:2005uc,Shih:2005qf,Jatkar:2005bh, Dabholkar:2006xa,David:2006ji,David:2006ru,David:2006yn,David:2006ud,Sen:2008ta,Sen:2009gy}. In an expansion in large charges it may be shown that this reproduces the correct semiclassical entropy of an extremal black hole carrying the same charges as the dyons. In many cases, higher-derivative and quantum corrections have also been computed on the macroscopic side and the results have been successfully matched with the corresponding corrections computed from the microscopic formula. We refer the reader to the reviews \cite{Sen:2007qy,Mandal:2010cj,Dabholkar:2012zz,Sen:2014aja} covering various aspects of this program for details and a more complete set of references. The computation of the quantum corrections is performed using the formalism of the Quantum Entropy Function \cite{Sen:2008vm,Sen:2008yk}. This proposal exploits the fact that the near--horizon geometry of extremal black holes always contains an $\adss$ factor \cite{Kunduri:2007vf,Figueras:2008qh}. In particular, for spherically symmetric black holes in four dimensions, the near--horizon geometry, embedded in 10--dimensional supergravity, contains an $\adss\otimes\ss2$ factor coupled to background $U(1)$ fluxes and scalar fields. The entire configuration is completely determined by the $SO(2,1)\otimes SO(3)$ isometry of the solution, along with the electric and magnetic charges carried by the black hole. In Euclidean signature, this configuration is given by
\begin{equation}\label{attractor}
\begin{split}
ds^2=a^2\l(d\eta^2+\sinh^2\eta d\theta^2\r) &+ a^2\l(d\psi^2 +\sin^2\psi d\phi^2\r), \quad 0\leq \eta<\infty,\, 0\leq\theta<2\pi,\\
F_{\eta\theta}^{(i)} = e^i, \quad F_{\psi\phi}^{(i)}&= {p_i\over 4\pi}\sin\psi,\quad  \Phi_w = u_w, \quad 1\leq i\leq r, \quad 1\leq w\leq s.
\end{split} 
\end{equation}
where the background has $r$ U(1) fluxes and $s$ scalar fields, and $a$ is a function of the electric and magnetic charges of the black hole, determined in terms of the $\l(e^i,p_i\r)$.

Using this fact it has been argued that the quantum degeneracy $d_{hor}\l(\vec{q}\r)$ associated with the horizon of an extremal black hole carrying charges $\vec{q}\equiv q_i$ is given by the unnormalized string path integral, with a Wilson line insertion, over all field configurations that asymptote to the attractor geometry of the black hole. In particular, \cite{Sen:2008vm,Sen:2008yk}
\begin{equation}\label{qef}
d_{hor}\l(\vec{q}\r)\equiv\l\langle\exp\l[i\oint q_i d\theta\mathcal{A}_\theta^i\r] \r\rangle_{\adss}^{finite}.
\end{equation}
The subscript `finite' reminds us that the path integral naively contains a volume divergence due to the presence of the $\adss$ factor. Regulating this divergence is carried out in accordance with the AdS/CFT correspondence. Though \eqref{qef} computes a degeneracy rather than an index, it may be shown that one may use this expression to compute the helicity trace index as well, which can then be compared with the microscopic results \cite{Sen:2009vz}.

Since its proposal, the conjecture of \cite{Sen:2008vm,Sen:2008yk} has been put to a variety of tests. Firstly, the leading saddle-point of the path integral is the attractor configuration \eqref{attractor} itself, and it may be shown that the value of the path integral \eqref{qef} at this saddle-point is the exponential of the Wald entropy associated with the black hole. Further, by expanding the massless fields of four--dimensional supergravity in quadratic fluctuations about this saddle-point, the logarithmic correction to the Wald entropy may be extracted from \eqref{qef} and matched with the microscopic answer \cite{Sen:2012dw}. This has been successfully carried out for the $1\over 4$--BPS black holes in $\cn=4$ supergravity and $1\over 8$--BPS black holes in $\cn=8$ supergravity \cite{Banerjee:2010qc,Banerjee:2011jp} and for rotating extremal black holes in \cite{Sen:2012cj}. The corresponding expressions for $1\over 2$--BPS black holes in $\cn=2$ supergravity have also now been obtained \cite{Sen:2011ba}, however in this case the microscopic results are so far not available. Recently, \cite{Keeler:2014bra} presented a new approach to the computation of logarithmic terms from \eqref{qef} which greatly simplifies the intermediate steps encountered in the calculations of \cite{Banerjee:2011jp,Sen:2011ba,Sen:2012cj}. We also note here that \eqref{qef} has been exactly evaluated for $\cn=4$ and $\cn=8$ string theories using localisation in \cite{Dabholkar:2010uh,Dabholkar:2011ec,Gupta:2012cy,Murthy:2013xpa,Dabholkar:2014ema} and the answer obtained precisely reproduces the microscopic expressions computed from the indices $B_n$.

Further, if we restrict ourselves to special subspaces of the moduli space which admit discrete symmetry transformations generated by an element $g$ and also require that the charges of the dyons be $g$--invariant, then we may define twisted indices as 
\begin{equation}
B^g_{2n} \equiv {1\over \l(2n\r)!}\tr\l[g\l(-1\r)^{2h}\l(2h\r)^{2n}\r].
\end{equation}
The group generated by $g$ is taken to be isomorphic to $\zn$. These indices were computed in \cite{Sen:2009md,Sen:2010ts}, and a proposal for their macroscopic interpretation was also presented in \cite{Sen:2009md}. In particular, \cite{Sen:2009md} considered Type II string theory compactified on $\mathcal{M}\otimes T^2$, where $\mathcal{M}$ could be either $T^4$ or $K3$, and $g$ was the generator of a geometric $\zn$ symmetry that acts on $\mathcal{M}$ and preserves 16 supercharges. The twisted index $B^g_6$, which receives contributions from dyonic states which preserve 4 supersymmetries all of which are $g$--invariant, was then computed. It was found that the answer in the large--charge limit takes the form \cite{Mandal:2010cj}
\begin{equation}
B^g_6\l(Q,P\r)=e^{\pi\sqrt{Q^2P^2-\l(Q\cdot P\r)^2}\over N}\l(\mathcal{O}\l(1\r)+\ldots\r).
\end{equation}
Therefore, if we assign an `entropy' to the index by taking its logarithm then we find that
\begin{equation}
\ln\vert B^g_6\l(Q,P\r)\vert = {S_{BH}\over N} +\mathcal{O}\l(1\r),
\end{equation}
i.e. the logarithmic correction to the entropy vanishes. Here
\begin{equation}
S_{BH}=\pi\sqrt{Q^2P^2-\l(Q\cdot P\r)^2},
\end{equation}
the Wald entropy of an extremal black hole carrying electric and magnetic charges $\l(Q,P\r)$. This is also the asymptotic expansion arrived at from Type IIB string theory on the CHL orbifold \cite{Sen:2010ts}. In this paper we shall show how this result arises from a macroscopic computation of the kind performed in \cite{Banerjee:2010qc,Banerjee:2011jp,Gupta:2013sva,Gupta:2014hxa} for the entropy of the black hole.

Before we do so, we briefly review the proposal made in \cite{Sen:2009md} regarding the macroscopic interpretation of the index $B^g_6$. The key ingredient of the proposal is that $B^g_6$ is indeed captured by a string path integral of the type \eqref{qef} in $\adss$. However, the path integral must now be carried out over fields which obey twisted boundary conditions along the $\theta$--circle of the $\adss$. In particular, as $\theta$ shifts by $2\pi$ the fields must transform by $g$. This partition function was denoted by $Z_g$ in \cite{Sen:2009md}. When we impose these boundary conditions then the attractor geometry itself is no longer an admissible saddle--point of the path integral as the $\theta$--circle is contractible in the interior of $\adss$, which leads to a singularity. Let us instead consider the following $\zn$ orbifold of the attractor geometry \eqref{attractor}, generated by the identification 
\begin{equation}\label{quotient}
\tilde{g}:\,\l(\theta,\phi\r)\mapsto\l(\theta+{2\pi\over N},\phi-{2\pi\over N}\r).
\end{equation}
Then it may be shown by an appropriate change of coordinates that the resulting field configuration still asymptotes to the full attractor geometry \eqref{attractor}. Additionally, this orbifold preserves enough supersymmetry that its contribution to the path integral \eqref{qef} does not automatically vanish by integration over the fermionic zero modes associated to broken supersymmetries. For these reasons, these field configurations are also admissible saddle--points of the quantum entropy function \eqref{qef}.\footnote{These orbifolds have fixed points at the origin of the $\adss$ times the north or south poles of $\ss2$ and \textit{a priori} it is not clear whether or not this is a consistent orbifold of string theory in the presence of background fluxes. If however the 10--dimensional attractor geometry also contains a circle $\mathcal{C}$ which is non--contractible at the origin of $\adss$, then one way to avoid this potential pitfall is to accompany the orbifold \eqref{quotient} by a translation by $1\over N$ units along $\mathcal{C}$. The orbifold group then acts freely over the 10--dimensional attractor geometry. If the radius of the circle $\mathcal{C}$ does not scale with the $\adss$ and $\ss2$ radii $a$, the precise details of the shift will not be relevant for us \cite{Sen:2012dw}. We do assume tacitly in our analysis that the generator $\tilde{g}$ includes such a shift along the internal directions as well. Such orbifolds have been explicitly defined in the 10--dimensional theory in \cite{Sen:2009vz,Banerjee:2009af}.} Using these inputs, \cite{Sen:2009md} proposed that $Z_g$ would receive contributions from the saddle--point obtained by imposing a $\zn$ orbifold generated by the action of $\tilde{g}$ on the attractor geometry, with $g$--twisted boundary conditions imposed on the fields. It was further shown that the value of $Z_g^{finite}$ at the saddle--point was given by $e^{S_{BH}\over N}$, in agreement with the asymptotic growth of $B^g_6$ from the microscopic side.

In this paper we will show that the correspondence between $Z_g$ and $B^g_6$ exists even at the quantum level. In particular, we will compute the log correction to the `entropy' given by $\log Z_g$ by expanding about the $\zn$ orbifold of the black hole attractor geometry generated by the action of $\tilde{g}$, where we impose $g$--twisted boundary conditions on the fields. We will find that the answer vanishes, in accordance with the microscopic results. In order to compute log corrections, we shall use the fact that the contributions of the form $\log\, a$ to the partition function of a theory defined with a length scale $a$ are completely determined from the one-loop fluctuations about the saddle-point, where we may focus exclusively on massless fields and further neglect higher-derivative terms \cite{Sen:2012dw}. Therefore the only fields that can contribute to the log term in $\log Z_g$ are the massless fields about its admissible saddle--points. We shall compute the log correction, focussing on modes which obey appropriate twisted boundary conditions, and find that the answer vanishes. While we do this computation explicitly for $\cn=8$ string theory obtained by compactifying Type II string theory on $T^6$, this is only for definiteness and we shall see that the results obtained would carry over to the $\cn=4$ case as well. We now give a brief overview of the computation, emphasizing the overall strategy and the important differences from the analyses previously carried out in \cite{Gupta:2013sva} and \cite{Gupta:2014hxa}. We will decompose the $\cn=8$ supergravity multiplet into irreducible representations of the $\cn=4$ subalgebra which commutes with $g$. These are one $\cn=4$ gravity multiplet, four $\cn=4$ gravitini multiplets and six $\cn=4$ vector multiplets, each of which are charged under $g$ as enumerated in Appendix \ref{charges}. Importantly for us the $\cn=4$ gravity multiplet is uncharged under $g$, and therefore obeys untwisted boundary conditions. Its contribution to the logarithmic term in the large charge expansion of $Z_g$ is therefore identical to that computed in \cite{Gupta:2014hxa}. The contributions of the gravitini and vector multiplets are however different from \cite{Gupta:2014hxa}, and are computed in this paper.

A brief overview of the paper is as follows. In section \ref{HK} we compute the heat kernel for scalars, Dirac fermions and `discrete modes' of the spin--1 and spin--$3\over 2$ fields on $\ads2s2n$ with twisted boundary conditions. This is an extension of the analysis of \cite{Gupta:2013sva} where the heat kernel over orbifold--invariant modes on these spaces was computed. We find that the answer again assembles into a global part, which obeys untwisted boundary conditions, plus conical contributions which are finite in the limit where the heat kernel time $t$ approaches zero. We put these results together to evaluate the contributions of $\cn=4$ vector and gravitino multiplets that obey twisted boundary conditions in section \ref{LC}. We find that the contribution to the log term vanishes for any non-zero value of the twist. These results demonstrate explicitly that the log term in $B^g_6$ vanishes for $\cn=8$ string theory and $\cn=4$ string theory. We then discuss how our results also prove that the log term vanishes even about exponentially suppressed corrections to the leading asymptotic formula for $B^g_6$ and conclude.
\section{The Heat Kernel for the Laplacian on $\ads2s2n$ with Twisted Boundary Conditions}\label{HK}
The goal of this paper is to compute logarithmic corrections to the partition function $Z_g$ defined as the path integral \eqref{qef} with $g$--twisted boundary conditions. These corrections only receive contributions from the one-loop fluctuations of massless fields over the $\zn$ orbifold of the attractor geometry generated by $\tilde{g}$. The one--loop partition function about this background is determined in terms of the determinant of the kinetic operator $D$ evaluated over the spectrum of the theory. We shall define this determinant by the means of the heat kernel method \cite{Vassilevich:2003xt}. The discussion below has has also been reviewed in the present context in \cite{Gupta:2013sva,Gupta:2014hxa} so we shall mainly recapitulate the key elements of the method.

We shall focus on operators of Laplace--type defined over fields on a manifold $\mathcal{M}$ with a length scale $a$. The eigenvalues of such operators scale as $1\over a^2$ and are denoted by $\kappa_n\over a^2$ and the corresponding degeneracies are $d_n$. With these inputs we may define the integrated heat kernel (referred from now on as simply `the heat kernel') as 
\begin{equation}\label{hkdef}
K\l(t\r) = \sum_{n} d_n e^{-{t\over a^2}\kappa_n}.
\end{equation}
Then the determinant of $D$ may be defined via
\begin{equation}\label{lndet}
-\ln\det D = \int_{\epsilon\over a^2}^\infty {d\bs\over \bs} K\l(\bs\r),
\end{equation}
where $\epsilon$ is a UV cutoff and $\bs={t\over a^2}$. Therefore, $\ln\det D$ contains a term proportional to $\ln a$, given by
\begin{equation}
-\ln\det D = 2 K_1 \ln a +\ldots,
\end{equation}
where $K_1$ is the $\mathcal{O}\l(\bs^0\r)$ term in the small $\bs$ expansion of the heat kernel $K(t)$ and the `$\ldots$' denote terms that are not of the form $\ln a$. From this expression, the term proportional to $\ln a$ in $\ln\mathcal{Z}$ may be extracted. Logarithmic corrections to black hole entropy have been computed from the quantum entropy function in this manner in \cite{Banerjee:2010qc,Banerjee:2011jp,Sen:2011ba,Sen:2012cj,Gupta:2013sva,Gupta:2014hxa}. We remind the reader that the small $\bs$ expansion of the heat kernel is in general non--trivial and contains ${1\over \bs^n}$ terms which have to be carefully computed. We will however find useful simplifications which enable us to analyze the problem efficiently.

Before proceeding further, we remind the reader that the analysis presented above has subtleties when the operator $D$ is only positive semi--definite, i.e. has zero modes. In that case the one--loop partition function contains the determinant of $D$ evaluated only over non--zero modes. The zero mode contribution needs to be analyzed separately \cite{Banerjee:2010qc,Banerjee:2011jp,Sen:2012dw,Bhattacharyya:2012ye}. The kinetic operator for which we compute the heat kernel is the one studied in \cite{Banerjee:2010qc,Banerjee:2011jp,Gupta:2013sva,Gupta:2014hxa}. This has zero modes over spin--2, spin--$3\over 2$ and spin--1 fields. However, the zero modes of the graviton and gravitino arise only within the $\cn=4$ gravity multiplet \cite{Banerjee:2011jp} which obeys untwisted boundary conditions in the path integral $Z_g$ and have therefore already been accounted in the analysis of \cite{Gupta:2014hxa}. Additionally, it may be shown that the log term for vectors may as well be extracted out by defining the heat kernel over all eigenvalues $\kappa_n$, including the zero eigenvalue, and extracting the $\mathcal{O}\l(\bs^0\r)$ term as before \cite{Banerjee:2010qc}. We will therefore ignore the presence of zero modes in our present analysis.

We now turn to the main computation of this section, which will provide us with the essential tools we need to compute logarithmic corrections to the partition function $\mathcal{Z}_g$. These are the heat kernels of the Laplacian over scalar fields and of the Dirac operator over spin-$1\over 2$ fields on $\ads2s2n$, where the $\zn$ orbifold is generated by $\tilde{g}$. The heat kernel over the fluctuations invariant under the $\tilde{g}$--generated $\zn$ orbifold was computed and the log term extracted in \cite{Gupta:2013sva,Gupta:2014hxa}. The analysis of this section is entirely analogous, with the only difference being that we now focus on modes which obey twisted boundary conditions under the $\tilde{g}$ orbifold. We find that the essential steps carry over directly from \cite{Gupta:2013sva,Gupta:2014hxa} with only minor modifications. For this reason, we shall focus on the scalar on $\ads2s2n$ to illustrate the steps and main modifications and then mostly enumerate final expressions for the spin-$1\over 2$ field. Further, as has been shown in \cite{Banerjee:2010qc,Banerjee:2011jp}, the higher--spin fields in the supergravity multiplets may be expanded in a basis obtained by acting on the scalar with the background metric and covariant derivatives and acting on the spin--$1\over 2$ field with gamma matrices and covariant derivatives. It turns out that the heat kernel over all quadratic fluctuations may be organised into the heat kernel over scalars and spin--$1\over 2$ fermions with appropriate multiplicities and shifts in eigenvalues. This will also be of great utility in our present analysis. Finally, we note that the heat kernel expression \eqref{hkdef} contains both eigenvalues and degeneracies of the kinetic operator $D$. On manifolds like $\adss$ the notion of degeneracy is subtle and requires a careful definition. It takes the form of the Plancherel measure \cite{Camporesi:1994ga,Camporesi:1995fb,Camporesi}. On quotients of AdS spaces, it turns out to be useful to exploit the fact that harmonic analysis on AdS is related to the sphere by an analytic continuation \cite{Camporesi:1994ga,Camporesi:1995fb,Camporesi}. By exploiting this analytic continuation, one may obtain the heat kernel and degeneracies of the Laplacian on these orbifolded spaces as well \cite{Gaberdiel:2010ar,Gopakumar:2011qs,Gupta:2013sva,Gupta:2014hxa}. We shall adopt this approach in this paper as well. In particular, we will consider the geometry given by
\begin{equation}
ds^2=a_1^2\l(d\chi^2 + \sin^2\chi d\theta^2\r)+a_2^2\l(d\psi^2 +\sin^2\psi d\phi^2\r),
\end{equation}
which is related via the analytic continuation 
\begin{equation}
\l(a_1,a_2\r)\mapsto\l(ia,a\r),\quad \chi\mapsto i\eta,
\end{equation}
to the $\ads2s2n$ geometry
\begin{equation}
ds^2=a^2\l(d\eta^2+\sinh^2\eta d\theta^2\r) + a^2\l(d\psi^2 +\sin^2\psi d\phi^2\r).
\end{equation}
The $\zn$ orbifold generated by $\tilde{g}$ acts on both these spaces via
\begin{equation}\label{orbact}
\tilde{g}:\quad \l(\theta,\phi\r)\mapsto\l(\theta+{2\pi\over N},\phi-{2\pi\over N}\r).
\end{equation}
Following the strategy of \cite{Gaberdiel:2010ar,Gopakumar:2011qs,Gupta:2013sva,Gupta:2014hxa}, we will do the computation on $\s2s2n$ and analytically continue the result to $\ads2s2n$. We will however need to be mindful of an important subtlety while performing this analytic continutation which arises due to a class of `discrete modes' of the vector and spin--$3\over 2$ fields in $\adss$ \cite{Camporesi:1994ga,Camporesi:1995fb}. These are normalisable eigenfunctions of the Laplacian over $\adss$ which are not related to normalisable eigenfunctions of the Laplacian over $\ss2$. Their contribution is computed separately in Section \ref{discretekernel}.
\subsection{The Heat Kernel for Scalars on $\ads2s2n$}\label{scalkernel}
In order to compute the heat kernel for the scalar Laplacian on $\ads2s2n$, we will first enumerate its spectrum \cite{Camporesi:1994ga}. The eigenvalues of the scalar Laplacian are  
\begin{equation}\label{scalareigenv}
E_{\lambda,\ell}={1\over a^2}\l(\lambda^2+{1\over 4}+\ell\l(\ell+1\r)\r),
\end{equation}
and the corresponding eigenfunctions are given by \cite{Camporesi:1994ga}
\begin{equation}\label{scalareigenf}
\Phi_{\lambda,\ell,m,n}\l(\eta,\theta,\psi,\phi\r) =f_{\lambda,m}\l(\eta,\theta\r)Y_{\ell,n}\l(\rho,\phi\r),
\end{equation}
where, omitting normalisation factors,
\begin{equation}\begin{split}
f_{\lambda,m}\l(\eta,\theta\r)=\l(\sinh^{\vert m\vert}\eta\r){_2F_1}\l(i\lambda+\vert m\vert+{1\over 2},-i\lambda\r.&+\l.\vert m\vert +{1\over 2},\vert m\vert +1,-\sinh^2{\eta\over 2}\r)e^{im\theta},\\& 0<\lambda<\infty,\quad m\in\mathbb{Z},
\end{split}\end{equation}
and the $Y_{\ell,n}$s are the usual spherical harmonics on $\ss2$. We will impose the projection \refb{orbact} generated by $\tilde{g}$ on the modes \eqref{scalareigenf} as in \cite{Gupta:2013sva}. The modes invariant under this orbifold are those for which $m-n=Np$, where $p$ is an integer. The heat kernel was computed over such modes in \cite{Gupta:2013sva}. We will look at the more general case for which 
\begin{equation}\label{cons1}
m-n =N p +q,\quad p\in \mathbb{Z}, \quad 0\leq q\leq N-1, \quad q\in\mathbb{Z}.
\end{equation}
We will refer to these as $q$-twisted boundary conditions. However, as mentioned above, we will carry out the computation by imposing the projection \eqref{orbact} on eigenfunctions of the scalar Laplacian on $\ss2\otimes\ss2$, which are given by 
\begin{equation}\label{s2scalef}
\Psi_{\tilde{\ell},m,\ell,n}\l(\chi,\theta,a_1,\rho,\phi,a_2\r)=Y_{\tilde{\ell},m}\l(\chi,\theta,a_1\r)Y_{\ell,n}\l(\rho,\phi,a_2\r).
\end{equation}
The corresponding eigenvalue is given by
\begin{equation}
E_{\tilde{\ell},\ell}= {1\over a_1^2}\tilde{\ell}\l(\tilde{\ell}+1\r)+{1\over a_2^2}\ell\l(\ell+1\r),
\end{equation}
which is related to $E_{\lambda\ell}$ by the analytic continuation
\begin{equation}\label{analytic}
\tilde{\ell}=i\lambda-{1\over 2},\quad \l(a_1,a_2\r)\mapsto \l(ia,a\r).
\end{equation}
Using the methods of \cite{Gupta:2013sva}, we find that the heat kernel on $q$--twisted modes on $\s2s2n$ is given by
\begin{equation}\label{kq}
K^q_s = {1\over N}K^s+{1\over N}\sum_{s=1}^{N-1}\sum_{\ell,\tilde{\ell}=0}^\infty\chi_{\ell,\tilde{\ell}}\l({\pi s\over N}\r) e^{-2\pi i qs\over N} e^{-t E_{\ell\tilde{\ell}}},
\end{equation}
where $K^s$ is the scalar heat kernel on the full unquotiented $\ss2\otimes\ss2$ space and the sum from $s=1$ to $N-1$ represents the contribution from the conical singularities and is expressed in terms of $\chi_{\ell,\tilde{\ell}}$, the $SU(2)\otimes SU(2)$ Weyl character
\begin{equation}
\chi_{\ell,\tilde{\ell}}\l({\pi s\over N}\r)\equiv \chi_{\ell}\l({\pi s\over N}\r)\chi_{\tilde{\ell}}\l({\pi s\over N}\r)\equiv {\sin{\l(2\ell+1\r)\pi s\over N}\over \sin\l({\pi s\over N}\r)}{\sin{\l(2\tilde{\ell}+1\r)\pi s\over N}\over \sin\l({\pi s\over N}\r)},
\end{equation}
where $\chi_{\ell}$ and $\chi_{\tilde{\ell}}$ are $SU(2)$ Weyl characters. The analytic continuation proceeds in the same way as for the untwisted case \cite{Gupta:2013sva,Gupta:2014hxa}. Firstly, the heat kernel over the unquotiented $\ss2\otimes\ss2$ gets continued to the heat kernel over $\adss\otimes\ss2$. Then the eigenvalue $E_{\tilde{\ell}\ell}$ gets continued to $E_{\lambda\ell}$ via \eqref{analytic}, and the Weyl character $\chi_{\tilde{\ell}}$ gets continued to the Harish--Chandra (global) character for $sl(2,R)$ \cite{Lang}
\begin{equation}
\chi^b_{\lambda}\l({\pi s\over N}\r)= {\cosh\l(\pi-{2\pi s\over N}\r)\lambda\over \cosh\l(\pi\lambda\r)\sin\l({\pi s\over N}\r)},
\end{equation}
and the conical terms get multiplied by an overall half \cite{Gupta:2013sva}. The factor of half accounts for the fact that under the $\zn$ orbifold \eqref{orbact}, $\adss\otimes\ss2$ has half the number of fixed points as does $\ss2\otimes\ss2$. Finally, the sum over $\tilde{\ell}$ gets continued to an integral over $\lambda$. We then obtain the heat kernel for the scalar on $\ads2s2n$ with the $q$-twisted boundary condition to be
\begin{equation}\label{ksqprimitive}
K^q_s= {1\over N}K^s + {1\over 2N}\sum_{s=1}^{N-1}\sum_{\ell=0}^{\infty}\int_0^\infty d\lambda \chi^b_{\lambda,\ell}\l({\pi s\over N}\r)e^{-{2\pi i qs\over N}}e^{-t E_{\lambda\ell}},
\end{equation}
where 
\begin{equation}
\chi^b_{\lambda,\ell}\l({\pi s\over N}\r)=\chi^b_{\lambda}\l({\pi s\over N}\r)\chi_{\ell}\l({\pi s\over N}\r).
\end{equation}
By doing the integral over $\lambda$ and the sum over $\ell$ as in \cite{Gupta:2014hxa} we find that \eqref{ksqprimitive} reduces to
\begin{equation}\label{ksq}
K_s^q={1\over N}K^s + {1\over 2N}\sum_{s=1}^{N-1}{1\over 4 \sin^4{\pi s\over N}}e^{-{2\pi i qs\over N}}+\ot.
\end{equation}
This is the expression we shall use to compute logarithmic corrections. It contains two terms. The first is the heat kernel of the untwisted scalar evaluated on the unquotiented space $\adss\otimes\ss2$. The second term is the contribution of the conical singularities. As observed in \cite{Gupta:2014hxa} for the untwisted modes, this term is finite in the limit where $t$ approaches zero. Hence the contribution of this term to the $\mathcal{O}\l(t^0\r)$ term in the heat kernel expansion is independent of the eigenvalue $E_{\lambda\ell}$. This will be of great utility in our further computations. Finally we note that the expressions \eqref{ksqprimitive} and \eqref{ksq} are divergent due to the infinite volume of $\adss$. However, using the prescription of \cite{Sen:2008vm,Sen:2008yk} this divergence may be regulated and a well--defined finite term extracted even on these quotient spaces \cite{Gupta:2013sva,Gupta:2014hxa}. Once this is done, we obtain a well--defined expression for the degeneracy $d^s_{\lambda\ell}$ of the eigenvalue $E_{\lambda\ell}$ in the $q$--twisted set of modes on $\ads2s2n$. This is given by
\begin{equation}\label{scalardeg}
d^s_{\lambda\ell}= -{1\over N}\l(\lambda\tanh\pi\lambda\r) \l(2\ell+1\r)+ {1\over 2N}\sum_{s=1}^{N-1} \chi^b_{\lambda,\ell}\l({\pi s\over N}\r)e^{-{2\pi i qs\over N}}.
\end{equation}
\subsection{The Heat Kernel for Fermions on $\ads2s2n$}
We will turn to the heat kernel of the Dirac operator evaluated over Dirac fermions on $\ads2s2n$ with $q$--twisted boundary conditions. The computations are entirely similar to those carried out in \cite{Gupta:2013sva,Gupta:2014hxa} once the $q$--twist has been accounted for as we have for the scalar in Section \ref{scalkernel}, we shall just mention the final result for the degeneracy of eigenvalues labelled by the quantum numbers $\lambda,\ell$ in the $q$--twisted set of modes on $\ads2s2n$.
\begin{equation}\label{fermiondeg}
d^f_{\lambda\ell}=-{8\over N}\l(\lambda\coth\pi\lambda\r) \l(\ell+1\r)+ {2\over N}\sum_{s=1}^{N-1} \chi^f_{\lambda,\ell+{1\over 2}}\l({\pi s\over N}\r)e^{-{2\pi iqs\over N}},
\end{equation}
where we have defined
\begin{equation}
\chi^f_{\lambda,\ell+{1\over 2}}\l({\pi s\over N}\r)=\chi^f_{\lambda}\l({\pi s\over N}\r)\chi_{\ell+{1\over 2}}\l({\pi s\over N}\r),
\end{equation}
and $\chi^f_\lambda$ is the Harish--Chandra character for $sl(2,R)$ given by \cite{Lang}
\begin{equation}
\chi^f_\lambda\l({\pi s\over N}\r)={\sinh\l(\pi-{2\pi s\over N}\r)\lambda\over \sinh\l(\pi\lambda\r)\sin\l({\pi s\over N}\r)}.
\end{equation}
We may use this degeneracy to obtain the heat kernel for the Dirac operator over the $q$--twisted Dirac fermions. We find that\footnote{We use the conventions of \cite{Banerjee:2010qc,Banerjee:2011jp} in which the fermion heat kernel is defined with an overall minus sign and is added to the bosonic heat kernels.}
\begin{equation}
K^q_f = {1\over N}K_f - {2\over N}\sum_{s=1}^{N-1}\sum_{\ell=0}^{\infty}\int_{0}^\infty d\lambda\, \chi^f_{\lambda,\ell+{1\over 2}}\l({\pi s\over N}\r)e^{-{2\pi iqs\over N}}e^{-t E_{\lambda\ell}}.
\end{equation}
As for the scalar, we may expand the conical term in a power series in $t$ omitting the $\ot$ and higher terms, carry out the $\lambda$ integral and the sum over $\ell$ to obtain
\begin{equation}\label{kfq}
K^q_f={1\over N}K_f-{1\over 2 N}\sum_{s=1}^{N-1}{\cos^2\l({\pi s\over N}\r)\over \sin^4\l({\pi s\over N}\r)}e^{-{2\pi i qs\over N}}+\ot.
\end{equation}
We will use \eqref{kfq} in our computations for the log term in Section \ref{LC}.
\subsection{The Heat Kernel over Discrete Modes}\label{discretekernel}
Vectors, gravitini and gravitons on the product space $\ads2s2n$ may be expanded in a basis contructed from the background metric, Gamma matrices and covariant derivatives, allowing us to express the heat kernel of the kinetic operator over supergravity fields in terms of the heat kernel over scalars and spin--$1\over 2$ fields \cite{Banerjee:2010qc,Banerjee:2011jp}. However, this analytic continuation fails to capture a set of discrete modes, labelled by a quantum number $\ell$, on the AdS space for the spin--1 and higher spin fields \cite{Camporesi:1994ga,Camporesi:1995fb,Camporesi}. The heat kernel over such modes needs to be computed directly on $\ads2s2n$.  Using the methods of \cite{Gupta:2013sva,Gupta:2014hxa}, we find that the degeneracy of an eigenvalue $E_{\ell}$ of the Laplacian over vector discrete modes obeying $q$--twisted boundary conditions is given by\footnote{We point out here that the modes with $\ell=0$ correspond to vector zero modes of the kinetic operator \cite{Banerjee:2010qc} and hence $d^{v_d}_{\ell=0}$ corresponds to the regularised number of vector zero modes of the kinetic operator. Explicitly evaluating \eqref{znonz} with $\ell=0$, so that $\chi_{\ell}\l({\pi s\over N}\r)=1$ $\forall s$, we find that $d^{v_d}_{\ell=0}$ vanishes when $q$--twisted boundary conditions are imposed. This is in contrast to the untwisted case, where $d^{v_d}_{\ell=0}=-1$\cite{Gupta:2013sva}.}
\begin{equation}\label{znonz}
d_\ell^{v_d}=-{2\ell+1\over N}-{1\over N}\sum_{s=1}^{N-1}\chi_{\ell}\l({\pi s \over N}\r)e^{-{2\pi i qs\over N}}.
\end{equation}
The degeneracy over the $q$--twisted gravitino discrete modes is given by
\begin{equation}\label{gravitinozero}
d_\ell^{f_d}= 8\l({\ell+1\over N}\r)-{4\over N}\sum_{s=1}^{N-1}{\sin{2\pi s\l(\ell+1\r)\over N}\over \sin{\pi s\over N}}\cos{\pi s\over N}e^{-{2\pi iqs\over N}}.
\end{equation}
Using the degeneracies \eqref{znonz} and \eqref{gravitinozero}, we can write down corresponding expressions for the heat kernels over these modes, though we do not do so explicitly here.
\section{Logarithmic Corrections to the Twisted Index}\label{LC}
We now turn to the computation of logarithmic corrections to $\mathcal{Z}_g$. We will carry out this computation for Type II string theory on $T^6$. This compactification preserves 32 supercharges of which 16 commute with $g$. Also, as we have previously discussed, the only fields which can contribute to the $\log a$ term are the massless fields in $\adss\otimes\ss2$. These are just the fields of four--dimensional $\cn=8$ supergravity. We will therefore find it useful to organise the spectrum of $\cn=8$ supergravity in terms of representations of the $\cn=4$ subalgebra which commutes with $g$. All the fields in a single $\cn=4$ multiplet are characterised by a common $g$-eigenvalue which in turn dictates which twisted modes on $\ads2s2n$ should the heat kernel be computed over. This information is summarised in Table \ref{t1}. In this section we shall compute the contribution of each multiplet in Table \ref{t1} to the log term in $\mathcal{Z}_g$, which requires us to compute the contribution to $\mathcal{Z}_g$ from quadratic fluctuations of massless fields about the $\zn$ orbifold generated by the action \eqref{orbact} of $\tilde{g}$ on the attractor geometry of the black hole. To do so, we shall compute the heat kernel of the kinetic operator derived in \cite{Banerjee:2010qc,Banerjee:2011jp} about this orbifolded background, imposing $g$--twisted boundary conditions on the fields as we act on the background with $\tilde{g}$. Therefore, the results of Section \ref{HK} will be useful for us.

Finally, as in \cite{Banerjee:2010qc,Banerjee:2011jp,Gupta:2013sva,Gupta:2014hxa}, we need to compute the heat kernel over the supergravity fields taking into account their couplings to the background graviphoton fluxes and scalar fields. As shown in \cite{Banerjee:2010qc,Banerjee:2011jp}, the heat kernel over the various quadratic fluctuations can be expressed in terms of the heat kernel over scalars, spin-$1\over 2$ fermions and discrete modes of higher--spin fields. The coupling to the background fields however changes the eigenvalues of the kinetic operator from those when fields are minimally coupled to background gravity. The new eigenvalues can in principle be computed by rediagonalising the kinetic operator. However, the flux does not change the degeneracy of the eigenvalue. Hence, to compute the heat kernel over the supergravity fields with our choice of background and boundary conditions, we can use the shifted eigenvalues computed in \cite{Banerjee:2010qc,Banerjee:2011jp} and the degeneracies computed in Section \ref{HK}. On doing so, we find two more simplifications that are of great benefit. Firstly, as observed in \cite{Gupta:2014hxa}, the contribution of the conical terms to the heat kernel is finite in the $t\mapsto 0$ limit. Hence the contribution to the  $\mathcal{O}\l(t^0\r)$ term from the conical terms is insensitive to the eigenvalues and can be computed from the degeneracies. Secondly, the other contribution to the $\mathcal{O}\l(t^0\r)$ term in the heat kernel originates from the $\mathcal{O}\l(t^0\r)$ term in the heat kernel computed for the full attractor geometry without imposing any twist on the boundary conditions. This has already been computed in \cite{Banerjee:2010qc,Banerjee:2011jp}. Using these results, and the $g$--charges computed in Table \ref{t1}, we can now compute the heat kernel over the various supergravity fields and extract the $\mathcal{O}\l(t^0\r)$ term in the heat kernel, which will yield the log term. With these results, we now turn to the main computation of this paper.

We firstly note that the $\cn=4$ gravity multiplet is $g$--invariant, and hence its heat kernel should be computed over untwisted modes. It has already been shown in \cite{Gupta:2013sva} that the contribution of these modes to the log term vanishes. Additionally, the contribution of any $g$--invariant $\cn=4$ vector multiplet to the log term also vanishes \cite{Gupta:2013sva}. Therefore we shall concentrate on the gravitino multiplets and the $\cn=4$ vector multiplets which carry a non--trivial $g$ charge, which corresponds to a non--zero twist in the boundary conditions. We find below that the contribution of these multiplets also vanishes for any arbitrary choice of twisting. This is in contrast to the untwisted case where while the contribution of the vector multiplet did vanish, the gravitino multiplet contribution was non-vanishing and was responsible for the non-zero log correction the entropy of $1\over 8$--BPS black holes in $\cn=8$ supergravity \cite{Gupta:2014hxa}.
\subsection{The Heat Kernel for the $\cn=4$ Vector Multiplet}\label{n4vec}
We will now put the results of Section \ref{HK} together, using the arguments presented above, to prove the first of our main results : the log correction in $\mathcal{Z}_g$ receives vanishing contribution from any $\cn=4$ vector multiplet with $q$--twisted boundary conditions. As in \cite{Banerjee:2010qc,Gupta:2013sva}, the heat kernel for any $\cn=4$ vector multiplet receives contributions from two Dirac fermions, 6 real scalars and one gauge field, along with two scalar ghosts. We will focus on the contribution of the conical terms to the $\mathcal{O}\l(t^0\r)$ term in the heat kernel. We denote this contribution by $K^c\l(t;0\r)$. Firstly the contribution from the two Dirac fermions is given by
\begin{equation}\label{vectmultf}
K^F_c\l(t;0\r)=-{1\over  N}\sum_{s=1}^{N-1}{\cos^2\l({\pi s\over N}\r)\over \sin^4\l({\pi s\over N}\r)}e^{-{2\pi i qs\over N}}.
\end{equation}
We now turn to the contribution from the integer--spin fields. These are the 6 real scalars, the gauge field and two scalar ghosts. Two of the scalars mix with the gauge field due to the graviphoton flux \cite{Banerjee:2010qc} and we have
\begin{equation}\label{kbdef1}
K^{B} = 4 K^s + K^{(v+2s)} - 2 K^s,
\end{equation}
where $K^s$ is the scalar heat kernel along $\ads2s2n$ with $q$--twisted boundary conditions, and $K^{(v+2s)}$ is the heat kernel of the mixed vector--scalar fields due to the background graviphoton flux. As we have previously argued, to extract the $t^0$ term from the fixed--point contribution to the heat kernel, we don't have to take into account the coupling of the gauge field to the scalars \textit{via} the graviphoton flux and can just add the various contributions piecewise. We therefore find that \eqref{kbdef1} reduces to
\begin{equation}\label{kbdef}
K^{B}_c=6K^s_c + K^v_c -2 K^s_c = 4K^s_c +K^v_c.
\end{equation}
$K^s_c$ can be read off from \eqref{ksq}, but we need to compute $K^v_c$. As shown in \cite{Banerjee:2010qc}, the heat kernel $K^v$ of a vector field over $\adss\otimes\ss2$ may be decomposed into $K^{\l(v,s\r)}$, which is the heat kernel of a vector field along $\adss$ times the heat kernel of a scalar along $\ss2$ and $K^{(s,v)}$, the heat kernel of a vector field along $\ss2$ times the heat kernel of a scalar along $\adss$. Further, the modes of the vector field along $\adss$ and $\ss2$ may be further decomposed into longitudinal and transverse modes. There is an additional discrete mode contribution from the vector field on $\adss$. These statements carry over to the case of the $\zn$ orbifolds with twisted boundary conditions as well. $K^v$ therefore receives the following contributions.
\begin{equation}\label{kvdef1}
K^v = K^{(v_T+v_L+v_d,s)}+K^{(s,v_T+v_L)}.
\end{equation}
Now the modes of longitudinal and transverse vector fields along $\adss$ and $\ss2$ are in one--to--one correspondence with the modes of the scalar with the only subtlety being that along $\ss2$ the $\ell=0$ mode of the scalar does not give rise to a non-trivial gauge field \cite{Banerjee:2010qc}. We therefore have
\begin{equation}
K^{(v_T,s)}= K^{(v_L,s)}=K^s,\quad K^{(s,v_T)}=K^{(s,v_L)}= K^s - K^{(s,\ell=0)},
\end{equation}
where, as we have mentioned previously, $K^s$ is the scalar heat kernel along $\ads2s2n$ with $q$--twisted boundary conditions, and $K^{(s,\ell=0)}$ is again the scalar heat kernel along $\ads2s2n$, however we only sum over the modes with $\ell=0$ along the $\ss2$ direction. We therefore find that the contribution of the conical terms\eqref{kvdef1} reduces to
\begin{equation}\label{kvdef}
K^v_c\l(t;0\r)=4K^s_c\l(t;0\r) + K^{(v_d,s)}_c\l(t;0\r) - 2 K^{(s,\ell=0)}_c\l(t;0\r).
\end{equation}
Further, using \eqref{scalardeg}, we may show that
\begin{equation}\label{kv1}
K^{(s,\ell=0)}_{c}\l(t;0\r)= {1\over 4N}\sum_{s=1}^{N-1}{1\over\sin^2\l({\pi s\over N}\r)}e^{-{2\pi i qs\over N}},
\end{equation}
and that $K^s_c\l(t;0\r)$ is given by
\begin{equation}\label{kv2}
K^s_c\l(t;0\r) = {1\over 2N}\sum_{s=1}^{N-1}{1\over 4 \sin^4{\pi s\over N}}e^{-{2\pi i qs\over N}}.
\end{equation}
Finally, using \eqref{znonz}, $K^{(v_d,s)}_c\l(t;0\r)$ is given by
\begin{equation}\label{kv3}
K^{(v_d,s)}_c\l(t;0\r)=-{1\over 2N}\sum_{s=1}^{N-1}{1\over \sin^2\l({\pi s\over N}\r)}e^{-{2\pi i qs\over N}}.
\end{equation}
Using \eqref{kbdef} and \eqref{kvdef}, and then putting \eqref{kv1}, \eqref{kv2} and \eqref{kv3} together, we find that the total integer--spin contribution is given by
\begin{equation}\label{vectmultb}
K^B_c\l(t;0\r) = {1\over N}\sum_{s=1}^{N-1}e^{-{2\pi i qs\over N}}\l({1-\sin^2\l({\pi s\over N}\r)\over \sin^4\l({\pi s\over N}\r)}\r).
\end{equation}
Then the total contribution of the conical terms from bosons and fermions is obtained by adding \eqref{vectmultf} and \eqref{vectmultb} to obtain
\begin{equation}\label{kvectmultc}
K^c\l(t;0\r)=K^B_c\l(t;0\r)+K^F_c\l(t;0\r) = {1\over N}\sum_{s=1}^{N-1}e^{-{2\pi i qs\over N}}\l({1-\sin^2\l({\pi s\over N}\r)-\cos^2\l({\pi s\over N}\r)\over \sin^4\l({\pi s\over N}\r)}\r)=0.
\end{equation}
This vanishes for arbitrary values of $q$. Now, using the arguments at the beginning of the section, the heat kernel for the $\cn=4$ vector multiplet about the $\tilde{g}$--generated $\zn$ orbifold of the attractor geometry is given, on imposing $q$--twisted boundary conditions, by
\begin{equation}
K^q = {1\over N}K + K^c\l(t;0\r) +\ot, 
\end{equation}
where $K$ is the heat kernel on the unquotiented near--horizon geometry. We therefore have, for the $t^0$ term in the heat kernel expansion,
\begin{equation}
K^q\l(t;0\r) = {1\over N}K\l(t;0\r) + K^c\l(t;0\r)
\end{equation}
We have shown in \ref{kvectmultc} that $K^c\l(t;0\r)$ equals zero. In addition, it was shown in \cite{Banerjee:2010qc} that $K\l(t;0\r)$ also vanishes. This implies that $K^q\l(t;0\r)$ also vanishes, which proves that the contribution to the log term from the vector multiplet vanishes even for $q$--twisted boundary conditions\footnote{We emphasize here that though the contribution of an $\cn=4$ vector multiplet vanishes for both twisted and untwisted boundary conditions, the origin of the result is different in both cases. For the untwisted case, the zero and non--zero modes of the kinetic operator give non--vanishing contributions to the log term which cancel against each other \cite{Gupta:2013sva}, while for the twisted case these contributions are individually zero as shown in this section and in footnote 3 of this paper.}. 
\subsection{The Heat Kernel for the $\cn=4$ Gravitino Multiplets}\label{n43half}
We now compute the contribution of the $\cn=4$ gravitino multiplets to the log term in $\mathcal{Z}_g$ for $\cn=8$ string theory. From Table \ref{t1}, we see that the $\cn=4$ gravitino multiplets obey $q$--twisted boundary conditions. There are four such multiplets, where the highest-weight field is a Majorana spin--${3\over 2}$ fermion, which we organise into two multiplets where the highest--weight field is a Dirac spin--$3\over 2$ fermion. One multiplet obeys twisted boundary conditions with $q=+1$, and the other with $q=-1$. Further, since we are considering quadratic fluctuations, the background flux in the attractor geometry does not cause gravitino multiplets with different $g$--charge, and hence different $q$--twist, to mix with each other. We will therefore focus on the contribution of the log term from one $q$--twisted multiplet where the highest--weight field is a Dirac spin--$3\over 2$ fermion. 

Now we shall compute the contribution of the conical terms to the $t^0$ term in the heat kernel expansion for this multiplet. Firstly, we focus on the integer--spin fields. There are 8 gauge fields and 16 real scalars. Further, gauge fixing introduces two ghost scalars for every gauge field. Hence the contribution of the integer--spin fields to the $\mathcal{O}\l(t^0\r)$ term from the conical terms in the heat kernel is
\begin{equation}
K^B_c\l(t;0\r) = 8K^v_c\l(t;0\r) + 16 K^s_c\l(t;0\r) -16 K^s_c\l(t;0\r) =8K^v_c\l(t;0\r),
\end{equation}
which therefore implies that
\begin{equation}\label{kbgravitinomult}
K^B_c\l(t;0\r) = {4\over N}\sum_{s=1}^{N-1}{1\over \sin^4{\pi s\over N}}e^{-{2\pi iqs\over N}} - {8\over N}\sum_{s=1}^{N-1}{1\over\sin^2{\pi s\over N}}e^{-{2\pi iqs\over N}}.
\end{equation}
We have used \eqref{kvdef} with \eqref{kv1}, \eqref{kv2} and \eqref{kv3} to arrive at this expression. We now turn to the contribution of the half--integer spin fields. We will focus on the contribution of one Dirac gravitino multiplet, which contains one Dirac gravitino and 7 Dirac spin-$1\over 2$ fields. The degrees of freedom reorganise themselves into in 4 Dirac fermions with $\ell\geq 0$, 6 Dirac fermions with only $\ell=0$ modes along the $\ss2$, 7 Dirac fermions with only $\ell\geq 1$ modes along the $\ss2$, one discrete Dirac fermion, and 3 ghost Dirac fermions \cite{Banerjee:2011jp,Gupta:2014hxa}. We can then show that
\begin{equation}
K^F_c\l(t;0\r) = 8K^f_c\l(t;0\r) - K^{(f,\ell=0)}_{c}\l(t;0\r) + K^{f_d}_c\l(t;0\r),
\end{equation}
where $K^f$ is the heat kernel for the Dirac fermion, $K^{(f,\ell=0)}$ is the heat kernel for the Dirac fermion with only $\ell=0$ modes along the $\ss2$ and $K^{f_d}$ is the heat kernel over one discrete Dirac fermion. Now
\begin{equation}
K^f_c\l(t;0\r) = -{1\over 2N}\sum_{s=1}^{N-1}{\cos^2{\pi s\over N}\over \sin^4{\pi s\over N}}e^{-{2\pi iqs\over N}},
\end{equation}
and 
\begin{equation}
K^{(f,\ell=0)}_{c}\l(t;0\r)= -{2\over N}\sum_{s=1}^{N-1}{\cos^2{\pi s\over N}\over \sin^2{\pi s\over N}}e^{-{2\pi iqs\over N}}.
\end{equation}
Further, using \eqref{gravitinozero}, we find that the discrete mode contribution from the conical terms is given by
\begin{equation}
K^{f_d}_c\l(t;0\r) = + {2\over N}\sum_{s=1}^{N-1}{\cos^2{\pi s\over N}\over \sin^2{\pi s\over N}}e^{-{2\pi iqs\over N}}.
\end{equation}
We finally obtain that the full half--integer spin contribution is given by
\begin{equation}\label{kfgravitinomult}
K^F_c\l(t;0\r) = -{4\over N}\sum_{s=1}^{N-1}{1\over \sin^4{\pi s\over N}}e^{-{2\pi iqs\over N}} +{8\over N}\sum_{s=1}^{N-1}{1\over \sin^4{\pi s\over N}}e^{-{2\pi iqs\over N}}-{4\over N}\sum_{s=1}^{N-1}e^{-{2\pi iqs\over N}}.
\end{equation}
Adding \eqref{kbgravitinomult} and \eqref{kfgravitinomult}, we find that the conical contribution to the $t^0$ term in the heat kernel for a given value of $q$ is
\begin{equation}
K_c\l(t;0\r)= -{4\over N}\sum_{s=1}^{N-1}e^{-{2\pi iqs\over N}}= +{4\over N},
\end{equation}
which is independent of $q$. Then the contribution of the $g$--twisted $\cn=4$ gravitino multiplets to the log term in $\mathcal{Z}_g$ is given by
\begin{equation}
K^g\l(t;0\r)={1\over N}K\l(t;0\r) + 2 K_c\l(t;0\r), 
\end{equation}
where $K\l(t;0\r)$ is the coefficient of the $t^0$ term in the heat kernel expansion of the gravitino multiplets about the unquotiented near--horizon geometry. This was computed to be $-8$ in \cite{Banerjee:2011jp}. We therefore find that $K^g\l(t;0\r)$ is given by
\begin{equation}\label{3halflog}
K^g\l(t;0\r)= -{8\over N} + {8\over N}=0.
\end{equation} 
Hence, the contribution of the $\cn=4$ gravitini multiplets to the logarithmic term in $\mathcal{Z}_g$ also vanishes. 
\subsection{The Zero Mode Analysis}
We will now take into account the presence of zero modes of the kinetic operator for $\cn=8$ supergravity fields expanded about the black hole near horizon geometry. The final result, as mentioned above, is that the zero mode analysis of \cite{Gupta:2014hxa} goes through unchanged, but since the zero mode analysis is an important part of the computation, we shall present the result explicitly. The following general result \cite{Banerjee:2010qc,Banerjee:2011jp}, see also \cite{Bhattacharyya:2012ye}, will be useful for us. Consider a theory with a length scale $a$ and fields $\phi_i$ such that the kinetic operator for quadratic fluctuations about a given background has $n^0_{\phi_i}\geq 0$ number of zero modes. Further, let the zero mode contribution to the path integral scale with $a$ as
\begin{equation}
\mathcal{Z}\simeq a^{n^0_{\phi_i}\beta_{\phi_i}}\mathcal{Z}_0,
\end{equation}
where $\mathcal{Z}_0$ does not scale with $a$, and the numbers $\beta_{\phi_i}$ have been explicitly determined in \cite{Banerjee:2011jp} for the vector field (see also \cite{Banerjee:2010qc}), the gravitino and the graviton. In particular
\begin{equation}
\beta_{v}=1,\quad \beta_{3\over 2} =3,\quad \beta_{g}= 2.
\end{equation}
In that case, the log term for the partition function is given by
\begin{equation}\label{logterm}
\ln\mathcal{Z}_{\log}= \l(K\l(0;t\r) + \sum_{\phi_i}n^0_{\phi_i}\l(\beta_{\phi_i}-1\r)\r)\ln a,
\end{equation}
where $K\l(0;t\r)$ is the coefficient of the $t^0$ term in the heat kernel expansion of the kinetic operator over of all fields $\phi_i$, evaluated on both zero and non--zero modes. Therefore, as far as the vector field is concerned, we may simply evaluate the heat kernel over all modes, extract the $t^0$ coefficient from there, and ignore zero modes. Further, for the $\cn=8$ kinetic operator, all the zero modes of the spin--$3\over 2$ and spin--2 fields are contained in the $\cn=4$ gravity multiplet \cite{Banerjee:2011jp}. This is quantised with untwisted boundary conditions and its contribution has already been evaluated on the orbifold space in \cite{Gupta:2014hxa}, where it was determined that 
\begin{equation}\label{nzero}
n_{3\over 2} = 2, \quad n_g = -2.
\end{equation} 
\subsection{Logarithmic Corrections to the Twisted Index}
Now we are in a position to put together the above results to show that the logarithmic corrections to the partition function $\mathcal{Z}_g$ vanish for the $\cn=8$ theory. To do so, we will need the coefficients $K\l(0;t\r)$ from the $\cn=4$ vector, gravitini and gravity multiplets, as well as the corresponding zero mode contributions. It has already been proven in \cite{Gupta:2013sva} that an untwisted $\cn=4$ vector multiplet has a vanishing contribution to the log term about our background. Further, we have seen in Section \ref{n4vec} that $K\l(0;t\r)$
for the $\cn=4$ vector multiplet with twisted boundary conditions vanishes, and in \eqref{3halflog} that
$K\l(0;t\r)$ for the $\cn=4$ gravitini multiplets with twisted boundary conditions also vanishes. Hence, the only non--vanishing contributions to $\ln\l(\mathcal{Z}_g\r)_{\log}$ come from the $\cn=4$ gravity multiplet, which obeys untwisted boundary conditions. For this multiplet (see Eq. 5.46 of \cite{Gupta:2014hxa})
\begin{equation}
K\l(0;t\r)=-2.
\end{equation}
Putting these results in \eqref{logterm} with \eqref{nzero}, we find that
\begin{equation}
\ln\l(\mathcal{Z}_g\r)_{\log}=0,
\end{equation}
which completes the proof that the logarithmic term in $\mathcal{Z}_g$ vanishes, in accordance with the microscopic results for $B^g_6$ for $\cn=8$ string theory.
\section{Conclusions}
In this paper we exploited the heat kernel techniques developed in \cite{Gupta:2013sva} to compute the logarithmic terms in the large charge expansion of the twisted index $B^g_6$ in $\cn=8$ string theory. These vanish, matching perfectly with the microscopic computation. Further, the result may be extended to the $\cn=4$ case as follows. Firstly, since $g$ commutes with all 16 supercharges in this case, we continue to classify fields into multiplets of the four--dimensional $\cn=4$ supersymmetry algebra. Secondly, we need to focus only on the massless supergravity fields over the near--horizon geometry as only these can contribute to the log term. Finally, the $g$ action on the various $\cn=4$ multiplets can be found out using techniques similar to the ones employed in the $\cn=8$ case. Since $g$ acts geometrically on the compact directions, the $\cn=4$ gravity multiplet still does not transform, and its contribution to the log term vanishes as per the analysis of \cite{Gupta:2014hxa}. The $\cn=4$ vector multiplets would carry non--trivial $g$--charges, corresponding to non--trivial $q$--twists for these fields in the path integral $\mathcal{Z}_g$. We have already seen that the contribution to the log term from $\cn=4$ vector multiplets vanishes for arbitrary twists $q$. Therefore, the log term vanishes even for $\cn=4$ string theory.

As a final observation, we note that the microscopic expression for $B^g_6$ contains exponentially suppressed corrections of the form
\begin{equation}
B^g_{6,p}\l(Q,P\r) \simeq e^{\pi\sqrt{Q^2P^2-\l(Q\cdot P\r)^2}\over Np}\l(\mathcal{O}\l(1\r)+\ldots\r),\quad p\in \mathbb{Z}_+,\, p\geq 2.
\end{equation}
Using the arguments of \cite{Gupta:2013sva} for the untwisted index we find that the logarithmic correction vanishes about these saddle--points as well. Following through the arguments of \cite{Sen:2009md}, a natural candidate for the macroscopic origin of these corrections corresponds to a saddle--point of $\mathcal{Z}_g$ obtained by taking a $\mathbb{Z}_{Np}$ orbifold of the attractor geometry, where again $g$--twisted boundary conditions should be imposed on the fields in the path integral. From the analysis presented in this paper, it follows that the log corrections to $\mathcal{Z}_g$ vanish about these saddle--points as well, which matches with the expectation from the microscopic side.
\acknowledgments
We would like to thank Justin David, Rajesh Gopakumar and especially Ashoke Sen for several very helpful discussions and correspondence. AC would like to thank ICTP, Trieste for hospitality while part of this work was carried out. SL would like to thank IACS, Kolkata and ICTS--TIFR, Bangalore for hospitality while part of this work was carried out. AC's work is supported by the DAE project 12--R$\&$D--HRI--5.02--0303. SL's work is supported by National Research Foundation of Korea grants 2005--0093843, 2010--220--C00003 and 2012K2A1A9055280. MS is supported by a J.C. Bose fellowship awarded to Ashoke Sen by the Department of Science and Technology, India. Finally, SL would like to dedicate this paper to the memory of Avijit Lal (14.12.1979 -- 10.04.2014) gratefully acknowledging his constant encouragement to pursue research in theoretical physics.
\appendix
\section*{Appendix}
\section{The $g$--Charges for the $\cn=8$ Supergravity Fields}\label{charges}
In this appendix we shall review how the $g$--twist acts on the fields of four--dimensional $\cn=8$ supergravity. As $g$ commutes with the $\cn=4$ subalgebra of the full $\cn=8$ algebra, we expect that the $\cn=8$ gravity multiplet will decompose into $\cn=4$ multiplets, each of which carry some charge under $g$. We shall obtain these charges by working with Type IIB supergravity compactified on $T^4\otimes T^2$ and studying the action of $g$ on the supergravity fields, which are the graviton $h_{MN}$, the two--form $B_{MN}$, the three--form flux $C_{MNP}$ and two 16--component Majorana--Weyl spinors.\footnote{The indices $M,N$ take values $0,\ldots,9$, while $\mu,\nu$ will take values $0,\ldots,3$ which label the non--compact directions. The indices $m,n$ will take values $4,\ldots,9$ and label the compact directions.}
This action of the $g$-twist on the Type IIB supergravity fields compactified on $T^4{\otimes}T^2$ can be realised in an appropriate complex coordinate system$(z^1,z^2)$ on $T^4$ and $(z^3)$ on $T^2$ as \cite{Sen:2007qy}.
\begin{equation}\label{geometric}
dz^{1}\rightarrow e^{\frac{2{\pi}i}{N}}dz^1 \hspace{1cm}
dz^{2}\rightarrow e^{-\frac{2{\pi}i}{N}}dz^2 \hspace{1cm}
d\bar{z}^{1}\rightarrow e^{-\frac{2{\pi}i}{N}}d\bar{z}^1 \hspace{1cm}
d\bar{z}^{2}\rightarrow e^{\frac{2{\pi}i}{N}}d\bar{z}^2
\end{equation}
\begin{equation}
dz^{3}\rightarrow dz^3 \hspace{1cm}
d\bar{z}^{3}\rightarrow d\bar{z}^3
\end{equation}
These transformations can be thought of as individual rotations along the two cycles of $T^4$. The $g$--action on the ten dimensional fields is realised as a field transformation under the different representations of the Lorentz group. In the four dimensional theory obtained on compactification, the $g$--action may be thought of as an internal symmetry. 

The compactification of the $\cn=2$ supergravity fields on $T^4{\otimes}T^2$ gives one $\cn=8$ gravity multiplet in 4 dimensions. This contains one graviton $h_{\mu\nu}$, 8 spin--$3\over 2$ Majorana fields, 28 spin--1 fields, 56 spin--$1\over 2$ Majorana fields and 70 real scalars.
\begin{table} { 
\begin{center}\def\st{\vrule height 3ex width 0ex}
\begin{tabular}{|c|c|c|c|c|c|c|c|c|c|c|} \hline 
Multiplet  & Number of Multiplets  & $g$--Eigenvalue 
\st\\[1ex] \hline \hline
Gravity & 1  & 1 \st\\[1ex] \hline
Gravitino &   2 & $e^{-{2\pi i\over N}}$  \st\\[1ex] \hline
Gravitino &   2 & $e^{{2\pi i\over N}}$   \st\\[1ex] \hline
Vector   &  4 & 1 \st\\[1ex] \hline
Vector        & 1 & $e^{-{4\pi i\over N}}$ \st\\[1ex] \hline
Vector & 1 & $e^{{4\pi i\over N}}$ 
\st\\[1ex] \hline \hline 
\end{tabular}
\caption{$g$--Charges of the $\cal{N}$=4 multiplets. It is natural to expect the gravity multiplet to remain invariant since the 4D spacetime metric $h_{\mu\nu}$ is a spacetime field and is unaffected by coordinate transformations on the internal directions.} \label{t1}
\end{center} }
\end{table}
The spin--2 field $h_{\mu\nu}$ is just the spacetime metric. The spin--1 fields come from $G_{{\mu}m}$, $B_{{\mu}m}$, $C_{mn\mu}$ and $A_{\mu}$. The scalars come from $G_{mn}$, $B_{mn}$, $A_{m}$, $C_{mnp}$, dualizing the components $C_{m\mu\nu}$ of the three--form field, and the axion and the dilation. The origin of the 8 spin--$3\over 2$ fields and 48 spin--$1\over 2$ fields lie in the spin--$3\over 2$ $\psi^{\alpha}_{\mu}$ and spin-1/2 $\varphi^{\alpha}_m$ multiplets obtained on compactification of the two 16 component Majorana--Weyl spinors over $T^6$. 8 of the remaining spin--$1\over 2$ fields come from the compactification of the two ten-dimensional $\psi^{\alpha}_{[10]}$ spinors.

The $g$--twist commutes with 16 of the 32 supersymmetries. Hence we split the $\cn=8$ gravity multiplet into one $\cn=4$ gravity multiplet, four gravitino, and six vector multiplets. All the members of a given $\cn=4$ multiplet carry the same $g$--charge since $g$ commutes with the $\cn=4$ subalgebra. The g-charge of every field has been found to conform with the g-charge of the multiplet it belongs to. The final results of this computation have been summarised in Table \ref{t1}.

\providecommand{\href}[2]{#2}\begingroup\raggedright\endgroup
\end{document}